\documentclass[twocolumn,tighten]{aastex61}
\usepackage{amsmath}
\usepackage{amssymb}
\usepackage{graphicx}

\submitjournal{ApJ}

\shorttitle{Multi-frequency scatter broadening evolution of pulsars}
\shortauthors{Krishnakumar et al.}

\begin{document}

\title{Multi-frequency scatter broadening evolution of pulsars - I}

\correspondingauthor{M.A. Krishnakumar}
\email{kkma@ncra.tifr.res.in}

\author{M.A. Krishnakumar}
\affiliation{Radio Astronomy Centre, NCRA-TIFR, Udagamandalam, India}
\affiliation{National Centre for Radio Astrophysics, Tata Institute of Fundamental Research, Pune, India}
\affiliation{Bharatiar University, Coimbatore, India}

\author{B.C. Joshi}
\affiliation{National Centre for Radio Astrophysics, Tata Institute of Fundamental Research, Pune, India}

\author{P.K. Manoharan}
\affiliation{Radio Astronomy Centre, NCRA-TIFR, Udagamandalam, India}
\affiliation{National Centre for Radio Astrophysics, Tata Institute of Fundamental Research, Pune, India}

\begin{abstract}

We present multi-wavelength scatter broadening observations of 47 pulsars, made with the Giant Metre-wave
Radio Telescope (GMRT), Ooty Radio Telescope (ORT) and Long Wavelength Array (LWA). The GMRT observations
have been made in the phased array mode at 148, 234, and 610 MHz and the ORT observations at 327 MHz. The 
LWA data sets have been obtained from the LWA pulsar data archive. The broadening of each pulsar as a
function of observing frequency provides the frequency scaling index, $\alpha$. The estimations of $\alpha$ 
have been obtained for 39 pulsars, which include entirely new estimates for 31 pulsars. This study increases
the total sample of pulsars available with $\alpha$ estimates by $\sim$50\%. The overall distribution of 
$\alpha$ with the dispersion measure (DM) of pulsar shows interesting variations, which are consistent with
the earlier studies. However, for a given value of DM a range of $\alpha$ values are observed, indicating 
the characteristic turbulence along each line of sight. For each pulsar, the estimated level of turbulence,
$C^{2}_{n_e}$, has also been compared with $\alpha$ and DM. Additionally, we compare the distribution of 
$\alpha$ with the theoretically predicated model to infer the general characteristics of the ionized 
interstellar medium (ISM). Nearly 65\% of the pulsars show a flatter index (i.e., $\alpha < 4.4$) than that 
is expected from the Kolmogorov turbulence model. Moreover, the group of pulsars having flatter index is
typically associated with an enhanced value of $C^{2}_{n_e}$ than those with steeper index.

\end{abstract}


\section{Introduction} \label{intro}

The effects of the ionised Interstellar Medium (ISM) on broadband pulsar signals can be studied 
extensively using radio frequency observations. One such effect is interstellar scattering, due to which 
the emitted pulse gets broadened in time (hereafter referred to as pulse scatter-broadening). Another effect
of interstellar scattering is the angular broadening of compact radio sources. While estimation of the
angular broadening of pulsar demands Very Large Baseline Interferometry (VLBI) observations, as pulsars are
point-like sources, pulse scatter-broadening can be estimated from temporal observations using a sensitive
radio telescope. In this study, we concentrate on the measurements of temporal broadening of pulsar
signal at multiple wavelengths.

Interstellar scattering is induced by the fluctuations in the free electron density in the ISM, which causes
the pulsed signal to travel along multiple paths. This broadens an otherwise sharp pulse in time with a
characteristic exponential decay time, $\tau_{sc}$, the scatter broadening time, as shown by \citet{w72}.
Pulse scatter broadening is more prominent for distant pulsars at low frequencies and also leads to a
diffraction pattern at the observer's plane, that decorrelates over a characteristic bandwidth $\delta\nu_d$,
such that 2$\pi\tau_{sc} \delta\nu_d = C_1$. The constant $C_1$ is expected to be of the order of unity
for a Kolmogorov-type turbulence \citep{cwb85}. However, \citet{lw2} attempted to address this problem with
the observation of two pulsars, and found it to be much higher than unity. For a reliable estimate of $C_1$,
one requires simultaneous measurement of both $\tau_{sc}$ and $\delta\nu_d$ at a given observing frequency.

The study by \citet{rick77} showed that, for an isotropic homogeneous turbulent medium, the scattering 
strength can be attributed to a power law spectrum of electron density as $P_{n_e}(q) = C^{2}_{n_e} 
q^{-\beta}$, where $C^{2}_{n_e}$ is the scattering strength, $q$ is the corresponding three-dimensional wave
number and $\beta$ is the spectral index ( 2 $< \beta <$ 4). The scattering time $\tau_{sc}$, increases with
decreasing observing frequency and is related to $\beta$ via $\alpha = 2\beta/(\beta-2)$, where $\alpha$ is
the frequency scaling index of $\tau_{sc}$. If the same volume of electrons is considered to be responsible
for dispersion and scattering, a simple scaling relation of $\tau_{sc} \propto \nu^{-4}$ DM$^{2}$ can be
shown to exist \footnote{Dispersion Measure (DM) is the integrated column density of electrons in the line of
sight towards the pulsar}\citep{sch68}. For a medium following the Kolmogorov model turbulence, one
expects $\beta = 11/3$, and consequently the scaling relation changes as $\tau_{sc} \propto \nu^{-4.4}$ DM$^
{2.2}$ \citep{rom86}.

In order to measure $\alpha$ along the line of sight to a pulsar, one requires multi-frequency observations,
preferably carried out simultaneously or with near simultaneity to alleviate variations of pulse scatter
broadening with time. \citet{cwb85} made multiple frequency observations of five pulsars and then using
either $\tau_{sc}$ or $\delta\nu_d$ measurements found that all of them closely follow the Kolmogorov model.
\citet{lkmll01,lmgka04} presented multi-frequency measurements of $\tau_{sc}$ for a set of high and low DM
pulsars respectively, which indicated a lower value for $\alpha$ than that expected from Kolmogorov model for
higher DM pulsars. Later, a concerted effort by \citet{lw1, lw2, lw3} resulted in 68 measurements of $\alpha$
across a wide range of DMs, although many of the $\tau_{sc}$ measurements used by them were tens of years
apart. They noted a dip in the $\alpha$ vs DM curve around a DM of 100 pc cm$^{-3}$, which is similar to
the trend seen at higher DM (DM $>$ 500 pc cm$^{-3}$). However with a limited number of measurements, it is
difficult to establish the significance of such variations of $\alpha$ with DM and this demands an effort to
enlarge the sample of $\alpha$ measurements.

We recently conducted an exhaustive study of pulsar scatter broadening measurements with the Ooty Radio 
Telescope (ORT), situated at Udhagamandalam (Ooty), India\footnote{ see \citet{sw71} for more information
about ORT}, which yielded $\tau_{sc}$ measurements for 124 pulsars at 327 MHz \citep{kmnjm15}. In this large 
sample of pulsars, only eight has multi-frequency $\tau_{sc}$ measurements. This motivated us to take up
additional low frequency observations at 148, 234 and 610 MHz using the Giant Metre-wave Radio Telescope
(GMRT). We have also used some of the low frequency profiles available in the EPN database\hspace{-0.1cm}
\footnote{http://www.jb.man.ac.uk/research/pulsar/Resources/epn/browser.html} in this study. We present 
$\tau_{sc}$ estimates for 44 pulsars observed using both the ORT and the GMRT and three pulsars with the 
Long Wavelength Array (LWA) \citep{srb15}. For 39 pulsars, we have more than three frequency measurements 
out of which 31 are completely new, enabling us to estimate the frequency scaling index, $\alpha$ accurately.
Combining all the published $\alpha$ measurements from the literature, we analyse 99 pulsars in this study
and examine the dependence of scattering across various lines of sight.

\section{Observations and Data Reduction}\label{obs}

Observations of 45 pulsars were carried out using the GMRT situated near Pune, India (see \citet{sw91} for
more details about GMRT). Based on our previous observations with the ORT (KMNJM15), we identified a
sample of pulsars, for which the pulse scatter broadening dominates the estimated intrinsic pulse width. 
We therefore selected a subset of the above sample for which : (1) no previous multi-frequency measurements
are available, (2) the estimation of $\tau_{sc}$ has ambiguity due to considerably less scattering at 327 MHz
and hence lower frequency measurements are required to obtain reliable values of $\tau_{sc}$, since scatter
broadening is expected dominate the intrinsic pulse shape and (3) the systematic error due to intrinsic
profile shape evolution as a function of frequency is likely to be not significant due to dominant scatter
broadening at lower frequencies. For these reasons, we have used both 234 MHz and 148 MHz extensively, but
for some high DM pulsars, we also made use of the 610 MHz band available at the GMRT for our observations.
The GMRT has a unique capability to cover a range of frequencies from 150 MHz to 1400 MHz and is a
particularly suitable instrument for such studies. The GMRT observations were conducted between 2015 July
9---August 1. We also used results from our previous observations with the ORT at 327 MHz as reported by
KMNJM15.

The GMRT observations were conducted using its phased array mode. During observations at each of the bands, 
we made sure to have at least 15 antennae in order to have adequate sensitivity. We chose all the antennae 
from the compact array in the central square of the telescope along with the nearest antenna in each arm. 
The short spacings were selected as the ionosphere affects the phase stability of the array at low 
frequencies and time-scale for this stability depends on baseline length. Phasing for our chosen array was
required at least every 30 mins at the lowest frequency. Observations at each band were conducted at
different epochs within a span of three weeks. The data were recorded using the GMRT Software Back-end (GSB)
available at the facility \citep{roy10}. Observations at 148 and 234 MHz were conducted using 256
spectral channels across a 16 MHz bandwidth, with a temporal resolution of 121 $\mu$s. At 610 MHz, the
observations were carried out using a 32 MHz band with 256 spectral channels and 121 $\mu$s time resolution.
The observations of each pulsar varied from 10 mins to 45 mins depending on its flux density and the
sensitivity of the synthesized phased array. The data thus obtained were processed further to remove strong
radio frequency interference (RFI) lines that were present in the band. Due to the presence of strong RFI at
low frequencies, some of the pulsars (three at 148 MHz and one at 234 MHz band) could not be detected. The
data were then converted to SIGPROC\footnote{www.sigproc.sourceforge.net} filterbank format for further
processing. The data were dedispersed to the nominal DM of the pulsar and folded with the topocentric period
{\bf using the polynomial coefficients generated by TEMPO2 that predict the period at a given epoch}\footnote
{http://www.atnf.csiro.au/research/pulsar/tempo2/index.php?n=Documentation.Predictive} generated by 
TEMPO2\footnote{www.atnf.csiro.au/research/pulsar/tempo2/} \citep{hem06,ehm06} by using the ephemeris 
available in the ATNF pulsar catalogue\footnote{http://www.atnf.csiro.au/people/pulsar/psrcat/}\citep{mht05}. 
The details of 327 MHz observations with the ORT are available in KMNJM15.

For three pulsars, we used the data from the LWA. The LWA covers a frequency range of 10$-$90 MHz
\citep{srb15}, where low DM pulsars ($DM \sim$ 10$-$50 pc cm$^{-3}$) are likely to show scatter-broadened
pulse profiles. We analysed the LWA data by dividing the full band in to 4.9 MHz bands on each tuning of
the LWA by using the public data available at LWA pulsar database\footnote{lda10g.alliance.unm.edu/PulsarArchive/}.

We have also made use of the profiles available at the EPN database at 410, 436, 610, 658 and 1420 MHz. 
These observations were taken several years back and estimates of $\tau_{sc}$ can therefore be affected by 
temporal variations of the inhomogeneities in the ISM. However, changes in $\tau_{sc}$ over observation 
epochs have not been reported for any of the pulsars under our study till now, except for Crab pulsar, 
B0531+21\citep{kuz08}. In any case, these observations were used for a small group of pulsars, where the 
estimated $\tau_{sc}$ near 327 MHz are in good agreement with our recent observations. Hence, we included 
these in our study.

\section{Data Analysis} \label{anal}

We performed analysis on the current dataset similar to that reported in KMNJM15. The observed pulse profile 
is a convolution of the intrinsic pulse shape $P_{i}(t)$ with the impulse response characterizing the pulse 
scatter broadening in the ISM $s(t)$, the dispersion smear across the narrow spectral channel $D(t)$
and the instrumental impulse response, $I(t)$. Following \citet{rm97} we have,

\begin{equation}
P(t) = P_{i}(t) \ast s(t) \ast D(t) \ast I(t),
\label{eq3}
\end{equation}

\noindent where $\ast$ denotes convolution. The rise time of the receivers and the back-end are {\bf small 
enough to neglect the effect} of $I(t)$, while $D(t)$ is a rectangular function of temporal width given by 
the dispersion smearing in the narrow spectral channel for incoherent dedispersion.

In Equation~\ref{eq3}, we are interested in the function $s(t)$ which gives the scatter broadening of the
pulse. \citet{lmgka04} had shown that the simple thin screen model fits the observed scattering very well. We
also follow the same analysis in our study for extracting the scatter broadening time-scale from the profile.
As shown by \citet{w72}, the function $s(t)$ representing the thin screen model can be expressed as

\begin{equation}
s(t)=\exp (-t/\tau_{sc})U(t),
\label{eq4}
\end{equation}

where $U(t)$ is a unit step function.

There are different methods in the literature that can be used to extract the $\tau_{sc}$.
\citet{lkmll01,lmgka04} and KMNJM15 used a high frequency profile as the template for fitting the
scatter broadened profile to estimate $\tau_{sc}$. \citet{bhat04} used a CLEAN based algorithm to extract
the unscattered intrinsic pulse shape and the scatter broadening, which does not require $apriori$ knowledge
about the intrinsic pulse shape. \citet{lw1, lw2, lw3} and \citet{rm97} followed a different method, where they used a
simple Gaussian (or multiple Gaussians in the case of multi-component profiles) as a template for fitting.
This method can lead to a systematic error in the estimates of $\tau_{sc}$ due to unknown pulse width,
particularly at high frequencies, where the effect of scattering is less dominant. Nevertheless, \citet{lw2}
have shown that such an approach, despite possibly even resulting in erroneous $\tau_{sc}$ estimates,
may not affect the frequency scaling index, $\alpha$. Recently, \citet{dem11} demonstrated another way
of extracting the $\tau_{sc}$ using the method of cyclic spectroscopy. Since this method preserves the phase
information of the signal, one can recover the intrinsic, unscattered pulse profile shape and also the 
impulse response of the ISM. This method requires huge amount of computing power and considering the fact
that recovering the intrinsic profile shape and consequently the $\tau_{sc}$ {\bf is poor for long period}
pulsars \citep{jon13}, we refrain from using this method. Hence, in the current study, we
follow the  method of KMNJM15, where we used an unscattered high frequency profile obtained from the EPN
database as the template. For those pulsars, where no high frequency profiles were available, we have made a
Gaussian profile with the high frequency pulse width obtained from the literature. This template profile, 
$P_{i}(t)$, was used, to obtain a best fit model by minimizing the normalized $\chi^{2}$ value defined by

\begin{equation}
\chi^2 = \frac{1}{(N-4) \sigma_{off}^{2}} \sum_{j=1}^{N}[P_{j}(t) - 
P_{i}(a,b,c,\tau_{sc})]^2, 
\label{eq5}
\end{equation}

\noindent where $\sigma_{off}^{2}$ is the off-pulse rms, $P_{j}(t)$ is the observed pulse profile, $P_{i}(t)$ 
is the model profile and $N$ is the total number of bins in the profile. The model profile $P_{i}$ is scaled 
with the pulse amplitude $a$, shifted by a constant offset $b$ in phase and fitted to a baseline $c$ to
minimize the $\chi^2$. For fitting purposes, we used the non-linear fitting routine ``mrqmin'' given in
Numerical Recipes \citep{press01}, where the errors in $\tau_{sc}$ are obtained from the covariance matrix.

After getting all the $\tau_{sc}$ estimates for a given pulsar at multiple frequencies, we have used a
straight line fitting algorithm in the log-log plane using the log $\tau_{sc} = -\alpha$ log $\nu + b$,
scaling the error bars properly (0.434 $\boldsymbol{\delta\tau_{sc} / \tau_{sc}}$) to get the $\alpha$.
We find that our results are in agreement with that we obtained from the fitting function
$\tau_{sc} = b \nu^{-\alpha}$ and are also consistent with those from the previous studies 
\citep{lmgka04, lw3}.

\subsection{Identification of the sources of errors and their characterization} \label{analsub}

The channel width of our observations with GMRT was 62.5 kHz at 148 and 234 MHz, whereas it was 130 kHz at 
610 MHz. At a DM of 200 pc cm$^{-3}$, this introduces a dispersion smear of 1, 8 and 32 ms at 610, 234 and 
148 MHz respectively. This can affect our $\tau_{sc}$ estimates. Our fitting procedure takes this into
account by convolving the template profile, $P_{i}(t)$, by a dispersion profile D(t), which is a rectangular
function of width equal to the dispersion smear at a given frequency for each pulsar, before fitting the
pulse scatter broadening function given by Equation \ref{eq4}. Another important factor that affects the 
$\tau_{sc}$ measurement is the evolution of the scatter broadening itself across the observing band. Since we
have used a bandwidth of 16 MHz at both 148 and 234 MHz, the evolution of scatter broadening from the top to
bottom of the band can be considerable. With a simple Kolmogorov model (frequency scaling index $\alpha$=4.4),
the pulse scatter broadening evolves by 1.35 times in the 16 MHz bandwidth at 234 MHz (226--242 MHz) and by
1.61 times at 148 MHz (140--156 MHz). It introduces a systematic error in most of our measurements, with a
relatively large bandwidth and small number of frequency channels.

To estimate the magnitude of the systematic error caused due to both the above effects, we have simulated 
pulse profiles at different DMs for several values of  $\tau_{sc}$, as calculated from the DM$-\tau_{sc}$ 
relation reported in KMNJM15. These profiles were generated by taking into account the dispersion smear and 
the evolution of $\tau_{sc}$ across the band at both 148 and 234 MHz. Typical noise was also added to the 
profiles to mimic real observations. The estimation of $\tau_{sc}$ for the total bandwidth was performed for 
the simulated profiles using our analysis procedure. Comparison of the assumed $\tau_{sc}$ for our
simulations and the estimated value suggests that if an increased error bar of 3$\sigma$ for the band centred
at 148 MHz and error bar of 2$\sigma$ for the band centred at 234 MHz is assumed, the expected $\tau_{sc}$ is
consistent with the estimated value. Hence, we have scaled the error bars appropriately at each of these
bands before estimating the frequency scaling index $\alpha$.

Since the error bars on some of the $\tau_{sc}$ values are considerably large at 148 and 234 MHz, a simple 
$\chi^{2}$-fit to the data set for estimating $\alpha$ is found to be insufficient. If the error bar is too 
large on a $\tau_{sc}$ measurement, the least squares fit will give less weightage to that particular point
and the fit will be dominated by the points having small error bars. This may lead to a considerable
uncertainty in the value of $\alpha$. To address this issue, we have performed Monte-Carlo simulations
to obtain the the reported $\alpha$ measurements in this paper. For each frequency, we generated 10000
normally distributed random numbers with our measured $\tau_{sc}$ as the mean and the error on $\tau_{sc}$ as
the standard deviation. A fit was performed considering each set of $\tau_{sc}$ at different frequencies to
find $\alpha$. The median of these $\alpha$ values is taken as the actual value and the error bars are
estimated from 5 and 95 percentiles. We found this method to be more reliable in estimating $\alpha$ and its
limits.

\section{Results} \label{result}

\begin{figure*}[ht]
\centering

\centering
\begin{minipage}{.5\linewidth}
    \includegraphics[width=\linewidth,angle=270.]{J1849-0636.ps}
\end{minipage}
\hfill
\begin{minipage}{.3\linewidth}
    \includegraphics[width=\linewidth]{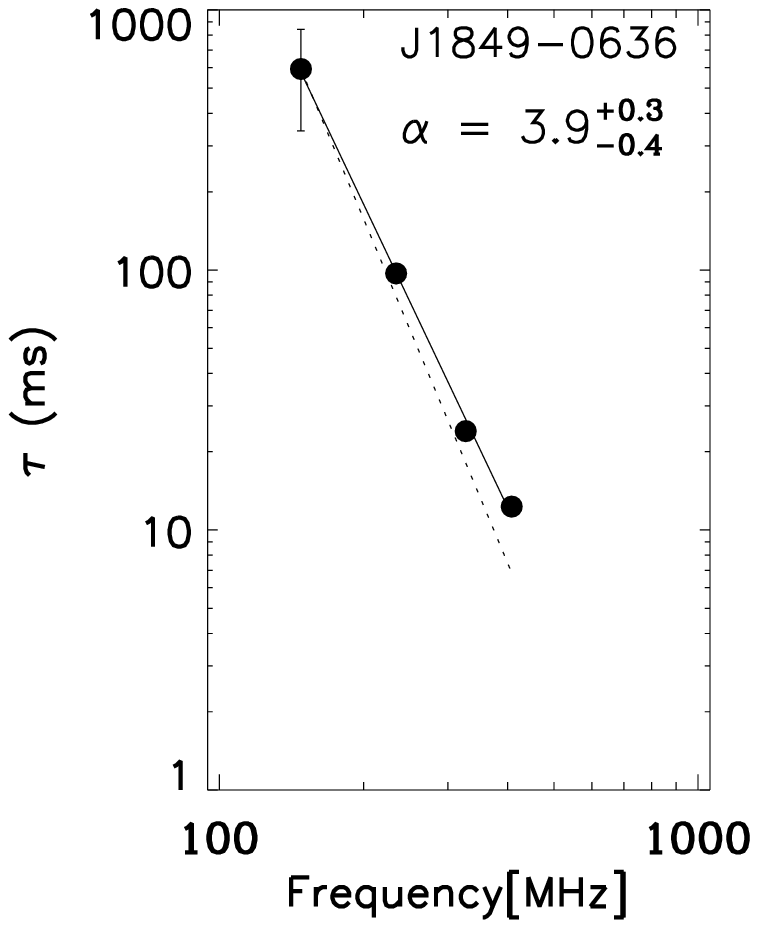}
\end{minipage}
\caption{{\bf Left panel}: The plot shows the frequency evolution of pulse scatter broadening as a function
of observing frequency for PSR J1849$-$0636. The profiles at 1420 and 408 MHz were taken from the EPN pulsar
database, whereas those at 327, 234 and 148 MHz were obtained from our observations. The black curve
represents the observed profile and the red curve represents the best fit model from which an estimate of 
$\tau_{sc}$ is obtained. {\bf Right panel}: The plot shows the fit to the $\tau_{sc}$ measurements (black
filled circles) as a function of observing frequency. The black continuous line shows the best fit model to
the dataset and the dashed line shows the expected line for the Kolmogorov turbulence model. The pulsar name
and the value of $\alpha$ obtained is also given in the right side of the panel along with confidence limits.}
\label{fig1}
\end{figure*}

The main aim of this work was to obtain new measurements of frequency scaling index $\alpha$, which required 
estimates of $\tau_{sc}$ at multiple frequencies. These were obtained from the fits as described in Section 
\ref{anal} for both the data acquired using the GMRT and archival data. The left panel of Figure~\ref{fig1} 
shows the frequency evolution of pulse scatter broadening as a function of observing frequency for PSR 
J1849$-$0636. As explained in KMNJM15, we have used the high frequency profile (at 1408 MHz in this case) as
the template for the fits shown by the red curves in the figure, which were used to estimate  $\tau_{sc}$ at
each frequency. The right panel of the Figure~\ref{fig1} shows the plot of the fit to the frequency evolution
of $\tau_{sc}$ and the estimated $\alpha$ values with error bars as detailed in Section~\ref{analsub}.

We are reporting  {\bf 25} new $\tau_{sc}$ measurements at 148 MHz, {\bf 41} at 234 MHz, {\bf 17} at 410 MHz 
(two are at 436 MHz) and {\bf eight} at 610 MHz (one is at 658 MHz). Out of these, 148 and 234 MHz
measurements are from our observations with the GMRT. At 410 and 610 MHz (except for five at 610 MHz which
are from our GMRT observations), the profiles were obtained from EPN data archive. All these measurements are
listed in Table~\ref{tab1}. In addition, we also obtained 20 estimates of $\tau_{sc}$ by analysing LWA
archival data \citep{srb15} at 35, 49, 64, 79 MHz tunings, by dividing the 19.6 MHz data into 4 sub
\hspace{-0.1cm}-bands, which are also listed in Table \ref{tab2}. 


The frequency scaling index $\alpha$ was estimated by fitting the measured $\tau_{sc}$ values with a power
law for those pulsars, where measurements for at least three frequencies were available. Due to this
reason, eight of the 44 pulsars in Table~\ref{tab1} did not make it to Table~\ref{tab3}. The fitted values
of $\alpha$ for all the {\bf 39} pulsars are reported\footnote{A complete set of plots of all the pulsars
with all details similar to that shown in Figure~\ref{fig1} which was used in estimating $\tau_{sc}$ for each
pulsar at available frequencies is reproduced in the supplementary data as well as at 
http://rac.ncra.tifr.res.in/data/pulsar/Supplementary-material-kbm17.pdf} in Table~\ref{tab3}. As {\bf 31} of
the $\alpha$ measurements are being reported for the first time, this has resulted in increasing the
currently available pool of $\alpha$ measurements by {\bf $\sim$50\%}. Table~\ref{tab3} summarises the 
results for the {\bf 39} pulsars in our study. For comparison with previous models, this table provides 
$\tau_{sc}$ scaled to 1 GHz, using the $\alpha$ estimated for each pulsar and also the $\tau_{sc}$
estimated from the NE2001 model \citep{cl02}. It can be seen that for most of the pulsars with $\alpha <$
4.0, NE2001 under estimates the $\tau_{sc}$.

In this study, we have expanded the pool of available multi-frequency $\tau_{sc}$ estimates by about 50\%.
Our new results are discussed in the next section in the context of previous studies.



\section{Multi-Wavelength evolution of scatter broadening} \label{discuss}

Out of the 39 measurements of $\alpha$ in our sample, eight pulsars were having previous measurements 
\citep{lmgka04, lw1, lw2}. Four of these measurements  were revised by \citet{lw3} and our new independent 
estimates of $\alpha$ are consistent with seven of those reported previously. They are, {\it viz} PSRs 
J0534+2200, J1935+1616, J2219+4755 \citep{lw2}, J1916+1312, J1932+2020, J2055+3630 \citep{lw3} and J2004+3137 
\citep{lmgka04,lw2,lw3}. Out of these, estimating $\alpha$ for PSR J0534+2200 is a difficult task. Variations 
in $\tau_{sc}$ are observed in this pulsar over weeks, due to the effects of ionised filaments in the Crab 
nebula crossing the line of sight. This can significantly affect $\alpha$, estimated from observations over 
different frequencies separated by a few weeks. The GMRT observations of the Crab pulsar at 148 and 234 MHz 
were separated by almost three weeks and the issue mentioned above can also affect our $\alpha$
estimate. To minimise such effects, we have taken simultaneous ORT observations on both of these days and
used them as the upper and lower limits at 327 MHz for fitting. This has resulted in a more robust estimate
of 3.4$\pm$0.2 for $\alpha$, which is within the error bars of the value quoted by \citet{lw3}. In the case
of PSR J1916+1312, our estimates are consistent with the revised values reported in \citet{lw2}.

Only one of our $\alpha$ measurements is different from what is reported earlier. This is in the case of PSR 
J0614+2229, where the earlier reported value was 1.7$\pm$0.5 \citep{lw2} and we obtain a value of 
2.9$\pm$0.1. \citet{lw2} used a single $\tau_{sc}$ measurement at 111 MHz with all other measurements at
and above 925 MHz. In contrast, our estimate is based on 148, 234, 327 and 408 MHz observations, with the
former two being near simultaneous measurements. Consequently, our estimate of $\alpha$ is likely to be more
reliable.

\begin{figure}
\includegraphics[scale=0.45,angle=-90]{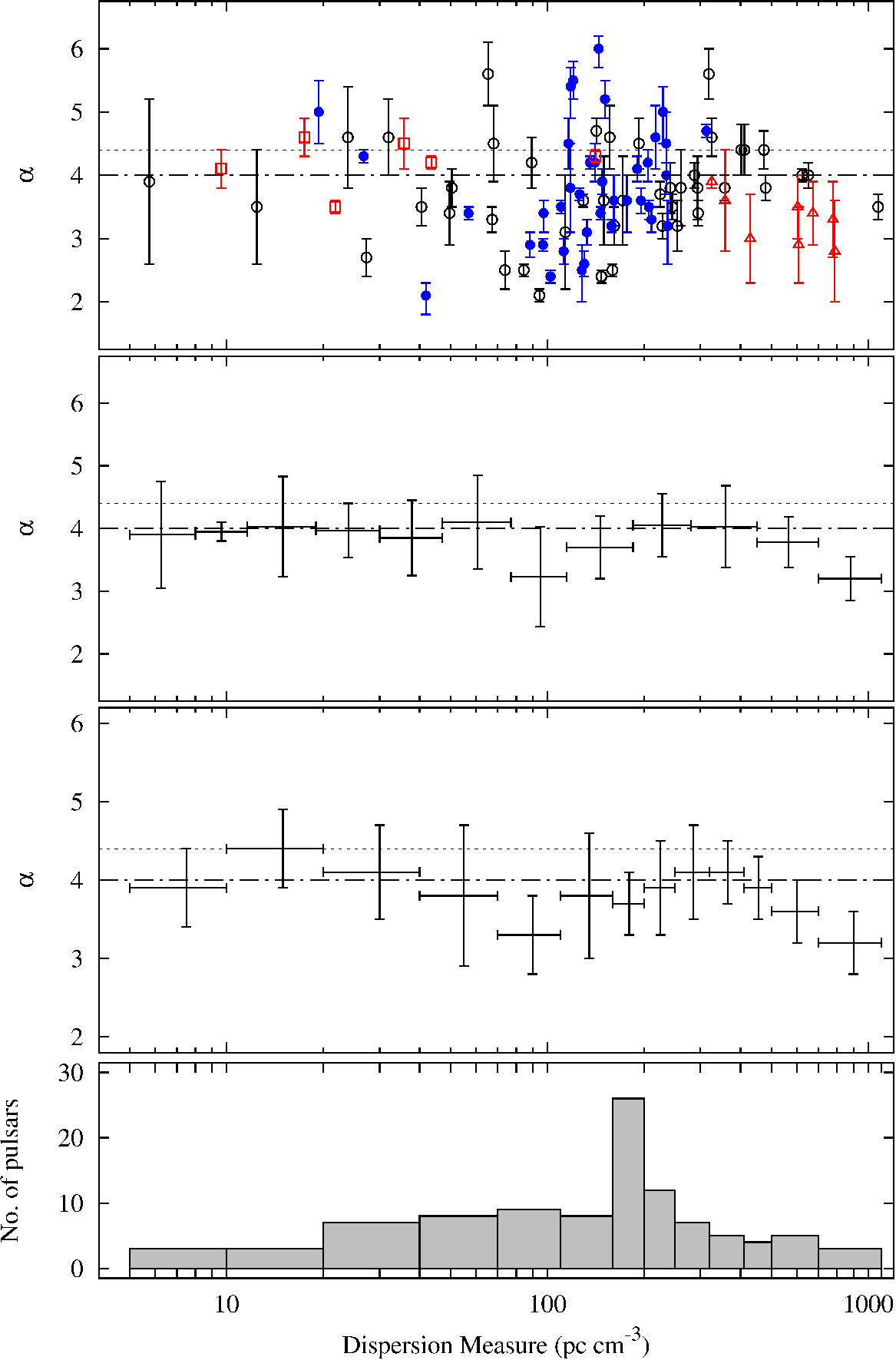}
\caption{{\bf Top panel}: Frequency scaling index, $\alpha$, of all the available pulsars are plotted against 
their DM. Black open circles are the measurements by \citet{lw1,lw2,lw3}, red open squares are the
measurements by \citet{cwb85,jnk98}, red filled triangles are the measurements by \citet{lkmll01, lmgka04}
and blue filled circles are our measurements. Dotted line corresponds to $\alpha$ = 4.4 and dash-dotted line
corresponds to $\alpha$=4.0. {\bf second panel}: Reproduction of $\alpha$ averaged over DM bins as shown by
\citet{lw3}. {\bf Third panel} $\alpha$ averaged over DM bins as explained in the text. {\bf Bottom panel}
Histogram showing the number of pulsars available in each of the DM bin.}
\label{fig3}
\end{figure}

All the available $\alpha$ measurements are plotted in the top panel of Figure~\ref{fig3} as a function of
DM. Second panel from the top shows the plot of $\alpha$ averaged over DM bins as shown in \citet{lw3}. Third 
panel from the top shows the average $\alpha$ over 13 DM bins, after including our new measurements. The 
averaging is done in such a way that at least 4 measurements are available at each of the DM bins. The bottom 
panel shows a histogram of the number of pulsars in each DM bin.

In order to understand the variation in $\alpha$ with DM, we divided the total DM range into four subsets. 
Subset $R1$ covers the low DM range from 0$-$50 pc cm$^{-3}$, $R2$ covers the mid-DM range from 50$-$250 pc 
cm$^{-3}$ and $R3$ covers the DM range from 250$-$500 pc cm${-3}$ and $R4$, the high DM range, i.e., DM above 
500 pc cm$^{-3}$. In the studies  so far, 13 $\alpha$ measurements are available in $R1$, 24 in the region 
$R2$ and a total of 24 in $R3$ and $R4$ \citep{lkmll01,lmgka04,lw1,lw2,lw3}. As it can be seen from Table 
\ref{tab3}, we are reporting 27 new and 7 updated measurements in $R2$, almost doubling the measurements in 
this range allowing us to better probe the existence of a transition DM or a dip in the $\alpha$--DM plane. 
In region $R1$, we obtained an average value of $\alpha$ as 3.9$\pm$0.5 from a total of 17 pulsars; in region 
$R2$, it is 3.7$\pm$0.8 from 58 pulsars; in region $R3$, it is 4.0$\pm$0.5 from a total of 16 pulsars and in 
region $R4$, the average $\alpha$ is 3.4$\pm$0.2 from 8 pulsars.

One of the implications of our measurements is an increase in the sample of available $\alpha$ by about 50\%. 
In particular, we have significantly increased the number of measurements in $R2$, where a departure from the 
value of $\alpha$ expected from a Kolmogorov spectrum was reported in the previous studies \citep{lkmll01, 
lmgka04}. While these authors reported a deviation at a transition DM of 300 pc cm$^{-3}$, \citet{lw1}
estimated transition DM to be 250 pc cm$^{-3}$. On the other hand, \citet{lw3} argued against any such
transition DM both at the mid-DM and high DM regime, based on the available sample of $\alpha$ measurements.
However, \citet{lw2,lw3} reported a noticeable dip in $\alpha-$DM relation at a DM of $\sim$100 pc cm$^{-3}$.
Based on their $\alpha$-DM relation, they concluded that $\alpha$ seems to be consistent with Kolmogorov
theory on an average for all DMs. Since all the pulsars in our study have DM $<$ 300 pc cm$^{-3}$, and the
fact that addition of new $\alpha$ measurements in this study did not show any noticeable departure from the
trend observed in $R3$, similar to what was shown by \citet{lw3} affirms that there is no possibility of a
transition DM in this range, as reported in \citet{lkmll01,lmgka04} and \citet{lw1}.

We further subdivided the pulsars in 13 DM bins and averaged the $\alpha$ for each bin as explained above.
Estimates of $\alpha$ are close to that expected from a Kolmogorov spectrum for pulsars with DM less than 50
pc cm$^{-3}$ and are consistent with previous results \citep{lkmll01,lmgka04,lw1,lw2,lw3}. However, these
studies suggest a marginal decrease in $\alpha$ at high DM regime. There is also a hint for the dip as
reported by \citet{lw2} in the DM range of 70$-$150 pc cm$^{-3}$ as it is evident from the Figure~\ref{fig3}.
Our measurements almost doubled the available measurements in this DM regime and one can see many low $\alpha$
values than what is expected in this range. Although there appears to be a transition in this DM range, which
is averaged over different lines of sight, one requires more measurements in this DM range to get a clear
understanding. The deviation of $\alpha$ from that expected for Kolmogorov spectrum seen at high DM ( $>$ 500
pc cm$^{-3}$) is similar to previous studies as our study does not add any significant measurements here.

Region $R1$ through $R3$ are quite interesting from the perspective of turbulence. In region $R1$, the
average value of $\alpha$ was 3.9$\pm$0.3. After adding our 4 new measurements in $R1$, the average value of 
$\alpha$ remained at 3.9$\pm$0.5. In $R2$ and $R3$, our new measurements are consistent with previous
similar studies \citep{lw2,lw3} and do not show remarkable change in the average $\alpha$ as is evident from
the Figure~\ref{fig3}. What is interesting is the fact that in region $R3$, the average $\alpha$ shows the
presence of a Kolmogorov type of turbulence. Though this has to be considered with caution, due to the low 
number statistics as well as an average over different lines of sight.

Nearly 50\% of pulsars in our sample have $\alpha$ flatter than what is expected from a homogeneous medium
with Kolmogorov model of turbulence. Seven of these pulsars have a distance of less than 3 kpc, while the
rest of the pulsars with a flatter $\alpha$ are all located beyond 3 kpc, where evidence for a departure 
from a Kolmogorov spectrum has also been reported (KMNJM15). About a third of our measurements are close to 
$\alpha$ = 4.4.

Based on these newly determined $\alpha$ measurements, we examined the relation between $\tau_{sc}$ and DM
after scaling $\tau_{sc}$ measurements to 1 GHz using the  $\alpha$ for 99 pulsars. These include 46
measurements from \citet{lw1,lw2,lw3}, eight from \citet{lkmll01,lmgka04}, six from \citet{cwb85} and 
\citet{jnk98} (some of these are based on scintillation bandwidth measurements) and 39 from the current
study. The general characteristic of this relation is consistent with that reported by \citet{lmgka04} and \citet{lw2}
with a flatter slope at low DM than that at higher DM.

\begin{figure*}[ht]
\centering
\hspace{-3.2 cm}
\includegraphics[scale=0.45,angle=-90.]{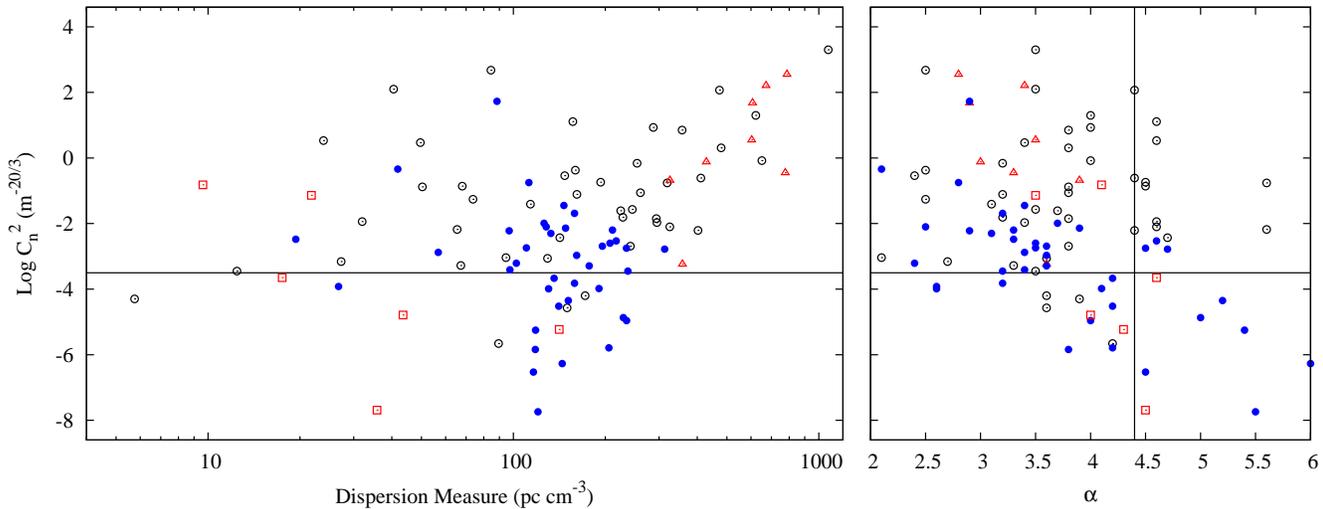}
\caption{{\bf Left panel}: The plot shows the distribution of log $C_{n_{e}}^{2}$ as a function of DM. The
black line shows the value of log $C_{n_{e}}^{2}$ = --3.5, expected for a Kolmogorov model turbulence. {\bf
Right panel}: The plot shows the distribution of the log $C_{n_{e}}^{2}$ as a function of $\alpha$. The black
vertical line at $\alpha$ = 4.4 is the expected value for a Kolmogorov model. The colour scheme is same as
that in Figure ~\ref{fig3}.
}
\label{fig4}
\end{figure*}

An interesting way of looking at this is by estimating the scattering strength, log $C_{n_{e}}^{2}$ along 
each line of sight and to study its distribution in different parameter spaces. The $C_{n_{e}}^{2}$ is 
estimated by using the relation $C_{n_e}^{2} = 0.002 \nu^{11/3} D^{-11/6} \delta \nu_d^{-5/6}$, where $\nu$
is the observing frequency in GHz, $D$ is distance to the pulsar in kpc and $\delta\nu_d$ is the
decorrelation bandwidth in MHz. A plot of the distribution of $C_{n_{e}}^{2}$ against DM and $\alpha$ is
given in Figure~\ref{fig4}. We observe a trend of increase in $C_{n_{e}}^{2}$ in relation with increase
in DM as we saw in KMNJM15. The right panel of the above figure is of more interest. We have divided the 
$\alpha-C_{n_{e}}^{2}$ plane into 4 quadrants, assuming Kolmogorov turbulence values for log $C_{n_{e}}^{2}$
= -- 3.5 and $\alpha$ = --4.4. If all the pulsars are passing through a medium which has fully developed
turbulence, we ought to see a cluster at the crossing point of the quadrants. This is not really the case as
one can see from the figure, although there is a small pack of pulsars close to the expected range.

Nearly 65\% of the pulsars are in the top-left quadrant, where the $\alpha$ is flatter and the log $C_{n_{e}}
^{2}$ is high. Interpretation of this trend requires some caution. We have assumed $C_1$=1.16, while
converting $\tau_{sc}$ to $\delta\nu_d$ using the relation 2$\pi\tau_{sc} \delta\nu_d = C_1$. While $C_1$ is
expected to be near to unity from theory, \citet{lw2} found $C_1$ to be 5 in the case of Vela pulsar. A value
of larger than the expected $C_{n_{e}}^{2}$ at $\alpha$ close to 4.4 strongly suggests that the value of $C_1$
is highly line of sight dependent. There are some pulsars in our sample, as evident from the 
Figure~\ref{fig4} whose $\alpha$ is indicative of a Kolmogorov turbulence model, but the log $C_{n_{e}}^{2}$
is very high. We abstain from scaling the log $C_{n_{e}}^{2}$ to --3.5 to estimate $C_1$ for these pulsars,
since it will result in very high value, which looks unrealistic. Unless one measures both $\tau_{sc}$ and 
$\delta\nu_d$ simultaneously at the same observing frequency, the determination of $C_1$ is difficult. The
above said scaling of log $C_{n_{e}}^{2}$ is applicable only when we have an independent estimate, in this
case $\alpha$, of the medium which shows that it follows Kolmogorov turbulence. For pulsars with a flatter 
$\alpha$, the above scaling will not be applicable. As it is clearly understood from the top-left quadrant of
Figure~\ref{fig4}, there is a tendency for the medium to have a flatter response in $\alpha$ and a larger
scattering strength. This require further modelling of the turbulence, as these lines of sight seem to
favour a non-Kolmogorov turbulence model.

\begin{figure*}[ht]
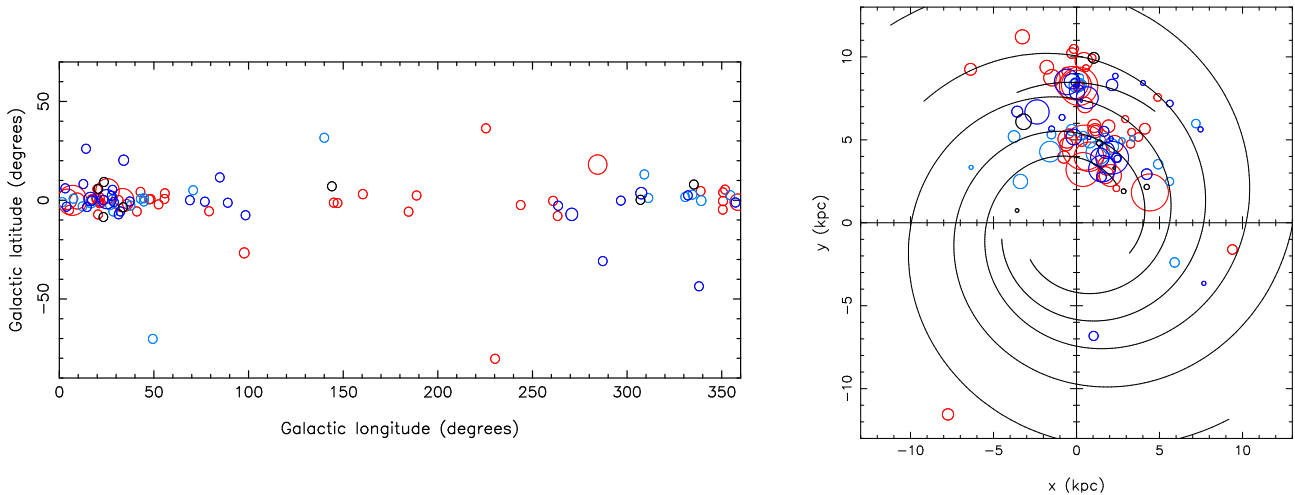

\centering

\centering
\begin{minipage}{.4\linewidth}
    \includegraphics[width=\linewidth,angle=270.,scale=0.7]{galLBC.ps}
\end{minipage}
\hfill
\begin{minipage}{.4\linewidth}
    \includegraphics[width=\linewidth,angle=270.,scale=0.9]{galXYC.ps}
\end{minipage}
\caption{The figure shows the position of the pulsars whose $\alpha$ is measured. Diameter of the circle
corresponds to respective $C_{n_{e}}^{2}$ values. Left panel shows the position of the pulsars in Galactic
longitude and latitude and the right panel shows the positions in galactic X-Y coordinates. Red circles
correspond to $\alpha$ values below 3.5, light blue circles correspond to $\alpha$ values between 3.5 -- 4,
dark blue circles represent $\alpha$ values between 4.0 -- 4.8 and black circles denote $\alpha$ values above
4.8. The spiral arm model of the Galaxy is taken from the YMW16 model \citep{ymw16}.}
\label{fig5}
\end{figure*}

In Figure~\ref{fig5}, we show the distribution of the pulsars for which the $\alpha$ values are known, in the
plane of the Galaxy. The left panel shows the $\alpha$ values as a function of its position in the Galactic
sky coordinates. The diameter of the circle is proportional to the magnitude of $C_{n{e}}^{2}$ of each
pulsar. The colour of the circles denote the range of $\alpha$ values. There is a cluster of blue and red
circles between longitude of 0 -- 50 degrees, as is seen in \citet{lw3}, probably because most of the
observed pulsars lie in this range. This cluster includes a range of $C_{n_{e}}^{2}$ values and corresponds
to different distances to pulsars as well as line of sight inhomogeneities. The right panel of Figure
\hspace{-0.2cm}~\ref{fig5} shows the position of the pulsars in the Galactic X-Y plane. The colour scheme is
the same as in the left panel. The distances are taken from the ATNF pulsar catalogue. In our sample, 38
pulsars have DM independent distance estimates. However, the distances to other pulsars have been obtained
from the YMW16 model \citep{ymw16} and they show considerable differences from those from NE2001 model. For
example, distance estimates from YMW16 model for pulsars J1623$-$4256 (X = -- 7.74, Y = -- 11.55),
J1807$-$2715 (X = 1.03, Y = -- 6.82) and J2305+3100 (X = 22.14, Y = 11.50, not shown in the figure, due to
scaling limitations) are not consistent with the ones from NE2001 model. In the case of YMW16 model, the
above pulsars are positioned farther than that from the NE2001 values. Two of them show shallower $\alpha$,
and low $C_{n_{e}}^{2}$ but J1807$-$2715 shows a near Kolmogorov turbulence characteristics. This clearly
shows that one needs to have DM independent distance estimates to pulsars to interpret our measurements shown
in Figure~\ref{fig5}.

The electron inhomogeneities can be very different at different lines of sight, such as those at Galactic
centre direction or at anti-centre direction and may imply different diffractive scales in comparison to the
inner scales for the Kolmogorov spectrum. It was shown by \citet{rick09} that an interplay of diffractive
scales and inner scale of Kolmogorov spectrum can explain $\alpha$ values smaller than 4.4. Thus averaging
over widely different lines of sight with very different electron density structures for a given DM bin can
produce trends of the type seen in Figures \ref{fig3} and \ref{fig5}. Hence, it is difficult to derive
meaningful interpretation about turbulence in the ISM given the scanty coverage of lines of sight in each DM
range with the present sample. We feel this question can better be addressed in the future by
increasing the current sample by 3 to 4 folds using new or upgraded telescopes like LWA1, LOFAR, upgraded
GMRT, etc. and by future telescopes such as Square Kilometre Array (SKA), FAST, etc.

The interplay between frequency dependence of diffractive scales and the inner scale of Kolmogorov turbulence
also implies that $\alpha$ can vary over different frequency ranges \citep{rick09}. Investigating this
requires measurements of $\tau_{sc}$ with fine frequency sampling, covering a wide range of frequencies, e.g.
200 $-$ 3000 MHz. Such observations have not been carried out to the best of our knowledge and the available
measurements (including those reported in this paper) sample sparsely the required frequency range.
Consequently, an estimate of $\alpha$ obtained from two frequency ranges which may correspond to different
scale sizes can result in an overall change in $\alpha$ that will be different from that predicted by
Kolmogorov turbulence.

\section{Summary and Conclusions}

In this paper, we present the frequency dependence of pulse broadening ($\alpha$) for a sample of 39 pulsars,
increasing the total available measurements by about 50\%. Out of the 39 pulsars, 36 were observed using the
ORT and GMRT at multiple frequencies and data for 3 pulsars were taken from the LWA pulsar database. We also
made use of the profiles at frequencies of 410 and 610 MHz in our study which were taken from EPN pulsar
database. With this, we studied the dependence of frequency scaling index of scatter-broadening ($\alpha$)
with DM. Our results are broadly consistent with other earlier studies. We find that $\alpha$ estimates for
DM less than 50 pc cm$^{-3}$ are in good agreement with those expected for Kolmogorov spectrum, but do notice
deviations beyond this DM as also reported by \citet{lmgka04} and \citet{lw3}. There is a large scatter in the data at
high DM range (DM $>$ 100 pc cm$^{-3}$). This could be due to multiple screens as mentioned in previous
studies, but can also be explained as a result of averaging over different lines of sight with different
diffraction scale size in comparison to inner scale of turbulence. Almost 65\% of the pulsars show a flatter
frequency dependence of scatter broadening evolution with a larger $C_{n_{e}}^{2}$, as compared to that
expected for a Kolmogorov turbulence model. This can get affected by the conversion factor $C_1$, which one
uses while converting $\tau_{sc}$ to $\delta\nu_{d}$. This requires further investigations, using
simultaneous measurements of both $\tau_{sc}$ and $\delta\nu_{d}$ at the same frequency.

The sample of $\alpha$ measurements need to be significantly enhanced to obtain a more uniform coverage of DM
in each line of sight to understand the turbulence characteristics better. This requires a concerted effort
to make multi-frequency measurements for a large sample of pulsars with a wide range of DMs. While this
is ideally possible with the three fold increase in pulsar population in the last decade, sensitive large
telescope with wide frequency coverage (and possibly multiple beams for commensal observations with other
pulsar projects to save observing time), such as upgraded GMRT, LOFAR, FAST, or SKA are required. These
telescopes will also provide a finer frequency sampling to investigate the inner scale effects, which may be
responsible for the features in $\alpha-$DM relations. Such observations with a finer frequency sampling are
planned in the near future with the currently available wide band back-ends at various telescopes.

{\it Acknowledgement}: We acknowledge the help and support provided by the observatory staff at both ORT and
GMRT. Both the facilities are operated and maintained by the National Centre for Radio Astrophysics of the
Tata Institute of Fundamental Research. We thank the LWA consortium, and the pulsar group at LWA for making
their data publicly available. We are grateful to Dipanjan Mitra for his helpful discussions and suggestions.
We are thankful to Yogesh Maan for his help in the LWA data reduction. We acknowledge support from Department
of Science and Technology grant DST-SERB Extra-mural grant EMR/2015/000515.

\startlongtable
\begin{deluxetable}{clcccccccccc}
\tablecolumns{12}
\tabletypesize{\footnotesize}
\tablecaption{Measurements of $\tau_{sc}$ for pulsars from our new observations with the GMRT, those reported
in KMNJM15  as well as archival data from EPN pulsar database. Each columns give pulsar name in J2000,
measured $\tau_{sc}$ and the $\chi^{2}$ of fit for 148, 234, 327, 410 and 610 MHz respectively. A `*' along
with the $\tau_{sc}$ measurement shows that the measurement is from 436 MHz profile and a `$\dagger$' denotes
that the measurement is from 658 MHz profile.
\label{tab1}}
\tablehead{
 & & \multicolumn{10}{c}{ $\tau_{sc}$ measurements} \\
\hline
No. & PSR & \multicolumn{2}{c}{148 MHz} & \multicolumn{2}{c}{234 MHz} & \multicolumn{2}{c}{327 MHz} & \multicolumn{2}{c}{410 MHz} & \multicolumn{2}{c}{610 MHz} \\
 & & $\tau_{sc}$ & $\chi^{2}$ & $\tau_{sc}$ & $\chi^{2}$ & $\tau_{sc}$ & $\chi^{2}$ & $\tau_{sc}$ & $\chi^{2}$ & $\tau_{sc}$ & $\chi^{2}$ \\
 & &   (ms)      &            &    (ms)     &            &   (ms)      &            &    (ms)     &            &    (ms)     &     \\
}

\startdata
1  & J0502+4654 & 102 $\pm$ 60 & 1.2 & 27 $\pm$ 5 & 0.9 & 19 $\pm$ 1 & 1.2 & 11 $\pm$ 1 \tablenotemark{a} & 0.9 &  - & - \\
2  & J0534+2200 & 25 $\pm$ 1 & 0.9 &	4.1 $\pm$ 0.2 & 1.0 & 2.25 $\pm$ 0.04 & 0.9 &  - & - & - & - \\
   & 		&		&	&		&	& 1.63 $\pm$ 0.01 & 0.9 &  &  &  & \\
3  & J0614+2229 & 23 $\pm$ 3 & 1.3 & 4.6 $\pm$ 0.6 & 0.7 & 1.74 $\pm$ 0.03 & 0.8 & 1.3 $\pm$ 0.1 \tablenotemark{a} & 1.2 &  - & - \\
4  & J1328$-$4921 & 19 $\pm$ 6 & 0.9 & 3.6 $\pm$ 0.5 & 1.2 & 0.8 $\pm$ 0.4 & 1.0 & - & - &	- & - \\
5  & J1557$-$4258 & - & - & 55 $\pm$ 18 & 0.8 & 5.3 $\pm$ 0.3 & 1.1 & 1.3 $\pm$ 0.1 \tablenotemark{b,*} & 1.0 & - & - \\
6  & J1604$-$4909 & 42 $\pm$ 29.0 & 0.7 & 7.2 $\pm$ 0.6 & 0.8 & 1.77 $\pm$ 0.04 & 0.6 & - & - & - & - \\
7  & J1613$-$4714 & 100 $\pm$ 70 & 0.7 & 23 $\pm$ 3 & 0.9 & 6.8 $\pm$ 0.5 & 0.9 & - & - & - & - \\
8  & J1639$-$4604 & - & - & 259 $\pm$ 78 & 1.1 & 26 $\pm$ 2 & 1.1 & - & - & - & - \\
9  & J1651$-$5222 & - & - & 29 $\pm$ 4 & 0.9 & 6.7 $\pm$ 0.4 & 1.1 & - & - & - & - \\
10 & J1703$-$3241 & 76 $\pm$ 8 & 0.8 & 15 $\pm$ 2 & 1.5 & 4.7 $\pm$ 0.1 & 0.9 & - & - & - & - \\
11 & J1705$-$3423 & - & - & - & - & 41 $\pm$ 3 & 0.9 & 23 $\pm$ 2 \tablenotemark{a} & 0.9 &  5.3 $\pm$ 0.1 & 1.0 \\
   & 		&   &   &   &   &            &     & 16 $\pm$ 1 \tablenotemark{c,*} & 1.1 &               &    \\
12 & J1722$-$3207 & - & - & 48 $\pm$ 2 & 0.9 & 13.9 $\pm$ 0.2 & 1.0 & 6.5 $\pm$ 1.4 \tablenotemark{a} & 0.8 & - & - \\
13 & J1731$-$4744 & 13 $\pm$ 1 & 1.0 & 4 $\pm$ 2 & 0.8 & 8.8 $\pm$ 0.2 & 0.9 & - & - & - & - \\
14 & J1732$-$4128 & - & - & - & - & 26 $\pm$ 4 & 1.2 & - & - & 0.4 $\pm$ 0.3 & 0.7 \\
15 & J1741$-$3927 & - & - & 39 $\pm$ 8 & 1.0 & 17.2 $\pm$ 0.4 & 1.0 & - & - & 1.8 $\pm$ 0.1 \tablenotemark{d,$\dagger$} & 0.8 \\
16 & J1743$-$1351 & 16 $\pm$ 3 & 0.8 & 2.3 $\pm$ 0.2 & 0.6 & 0.4 $\pm$ 0.1 & 1.0 &  - & - & - & - \\
17 & J1745$-$3040 & 39$\pm$ 14 & 0.7 & 12.1 $\pm$ 0.4 & 1.0 & 3.7 $\pm$ 0.2 & 1.1 & - & - & - & -\\
18 & J1759$-$2205 & 168 $\pm$ 126 & 1.3 & 44 $\pm$ 6 & 1.0 & 9.0 $\pm$ 0.3 & 1.0 & - & - & - & - \\
19 & J1801$-$0357 & 42 $\pm$ 7 & 0.9 & 2.0 $\pm$ 0.2 & 0.9 & 0.6 $\pm$ 0.1 & 1.0 &  - & - & - & - \\
20 & J1807$-$0847 & 38 $\pm$ 15 & 0.8 & 12 $\pm$ 1 & 0.6 & 4.3 $\pm$ 0.1 & 1.0 & 2.4 $\pm$ 0.2 \tablenotemark{a} & 0.9 & - & - \\
21 & J1807$-$2715 & - &  - & 151 $\pm$ 46 & 1.0 & 37 $\pm$ 2 & 1.3 & - & - & 1.8 $\pm$ 0.1 \tablenotemark{a} & 1.5 \\
22 & J1808$-$0813 & - & - & 85 $\pm$ 14 & 0.6 & 12 $\pm$ 1 & 0.8 & 4.4 $\pm$ 0.4 \tablenotemark{a} & 0.7 & - & - \\
23 & J1816$-$2650	& 55 $\pm$ 37 & 0.8 & 19 $\pm$ 6 & 1.1 & 7.5 $\pm$ 0.3 & 0.9 & - & - & - & - \\
24 & J1822$-$2256 & - & - & 66 $\pm$ 6 & 1.0 & 14.5 $\pm$ 0.2 & 1.0 & - & - & - & - \\
25 & J1823$-$0154 & - &  - & 44 $\pm$ 5 & 0.9 & 5.9 $\pm$ 0.3 & 1.2 & 4.8 $\pm$ 0.1 \tablenotemark{a} & 1.3 & - & - \\
26 & J1835$-$1106 & - & - & 25 $\pm$ 3 & 0.9 & 6.8 $\pm$ 0.4 & 1.0 & 4.5 $\pm$ 0.1 \tablenotemark{b} & 0.7 & 1.0 $\pm$ 0.1 & 1.0 \\
27 & J1849$-$0636	& 593 $\pm$ 250 & 1.0 & 97 $\pm$ 3 & 1.0 & 24 $\pm$ 1 & 1.3 & 12.3 $\pm$ 0.5 \tablenotemark{a} & 1.0 & - & - \\
28 & J1854+1050 & - & - & 138 $\pm$ 48 & 1.1 & 46 $\pm$ 4 & 1.3 & - & - & 4.9 $\pm$ 0.1 \tablenotemark{a} & 0.8 \\
29 & J1854$-$1421 & - & - & 7.3 $\pm$ 0.6 & 0.8 & 3.0 $\pm$ 0.2 & 1.0 & 1.7 $\pm$ 0.1 \tablenotemark{a} & 1.2 & - & - \\
30 & J1903$-$0632	& - & - & 66 $\pm$ 4 & 0.7 & 20.7 $\pm$ 0.3 & 0.9 & 8.3 $\pm$ 0.6 \tablenotemark{a} & 1.1 & - & -\\
31 & J1904$-$1224 & 397 $\pm$ 250 & 0.8 & 24 $\pm$ 5 & 1.0 & 5.6 $\pm$ 0.2 & 0.9 & - & - & - & - \\
32 & J1905$-$0056 & 380 $\pm$ 280 & 1.4 & 39 $\pm$ 3 & 0.6 & 7.4 $\pm$ 0.4 & 0.5 & - & - & - & - \\
33 & J1910$-$0309 & 75$\pm$ 33 & 1.1 &  16 $\pm$ 1 & 1.1 & 2.7 $\pm$ 0.1 & 0.9 & - & - & - & - \\
34 & J1910+0714 & - & - & 12 $\pm$ 2 & 0.8 & 4 $\pm$ 1 & 0.7 & - & - & - & - \\
35 & J1910+1231 & - & - & - & - & 52 $\pm$ 5 & 1.0 & - & - & 2.3 $\pm$ 0.2 & 0.6 \\
36 & J1916+1312 & 79 $\pm$ 60 & 0.8 & 24 $\pm$ 6 & 0.6 & 6 $\pm$ 1 & 1.0 & - & - & - & -\\
37 & J1926+0431 & 12 $\pm$ 1 & 0.7 & 4.0 $\pm$ 0.6 & 0.9 & 2.2 $\pm$ 0.1 & 0.9 & 0.9 $\pm$ 0.3 \tablenotemark{a} & 0.6 & - & - \\
38 & J1932+2020 & - & - & 71 $\pm$ 6 & 1.0 & 19 $\pm$ 1 & 1.1 & 12 $\pm$ 1 \tablenotemark{a} & 1.1 & - & - \\
39 & J1932+2220 & - & - & 7 $\pm$ 2 & 0.7 & 0.65 $\pm$ 0.03 & 1.1 & - & - & - & - \\
40 & J1935+1616 & 20 $\pm$ 2 & 1.0 & 4.5 $\pm$ 0.2 & 1.0 & 3.2 $\pm$ 0.2 & 0.7 & 1.2 $\pm$ 0.4 \tablenotemark{a} & 1.1 & 0.4 $\pm$ 0.1 \tablenotemark{a} & 0.9\\
41 & J2004+3137 & 222 $\pm$ 105 & 0.7 & 38 $\pm$ 2 & 1.1 & 9.0 $\pm$ 0.2 & 1.2 & - & - & - & -\\
42 & J2029+3744 & 295 $\pm$ 81 & 0.7 & 49 $\pm$ 3 & 1.0 & 11.6 $\pm$ 0.3 & 1.1 & - & - & - & -\\
43 & J2055+3630 & 85 $\pm$ 50 & 1.0 & 23 $\pm$ 2 & 0.8 & 5.5 $\pm$ 0.3 & 0.9 & 3.0 $\pm$ 0.1 \tablenotemark{a} & 0.9 & - & -\\
44 & J2113+4644 & 50 $\pm$ 2 & 1.3 & 5.6 $\pm$ 0.2 & 1.2 & 1.8 $\pm$ 0.1 & 0.7 & - & - & - & -\\
\enddata
\tablerefs{ : (a) -- \citet{gl98}; (b) -- \citet{dsb+98}; (c) -- \citet{lor94}; (d) -- \citet{mhq98}.}
\end{deluxetable}


\startlongtable
\begin{deluxetable}{cccccccccccccccccccc}
\tablecolumns{20}
\tablewidth{0pt}  
\tabletypesize{\tiny}
\tablecaption{Measurements of $\tau_{sc}$ for pulsars with archival data from LWA \citep{srb15}. Each column
gives pulsar name in J2000, measured $\tau_{sc}$ and the $\chi^{2}$ of fit for 35, 45, 50, 55, 65 and 80 MHz
respectively. \label{tab2}}

\tablehead{
 & & \multicolumn{18}{c}{$\tau_{sc}$ measurements} \\
\hline
No. & PSR & \multicolumn{2}{c}{37.55 MHz} & \multicolumn{2}{c}{42.45 MHz} & \multicolumn{2}{c}{47.35 MHz} & \multicolumn{2}{c}{52.27 MHz} & \multicolumn{2}{c}{57.15 MHz} & \multicolumn{2}{c}{62.05 MHz} & \multicolumn{2}{c}{66.95 MHz} & \multicolumn{2}{c}{71.85 MHz}  & \multicolumn{2}{c}{76.75 MHz} \\
 & & $\tau_{sc}$ & $\chi^{2}$ & $\tau_{sc}$ & $\chi^{2}$ & $\tau_{sc}$ & $\chi^{2}$ & $\tau_{sc}$ & $\chi^{2}$ & $\tau_{sc}$ & $\chi^{2}$ & $\tau_{sc}$ & $\chi^{2}$ & $\tau_{sc}$ & $\chi^{2}$ & $\tau_{sc}$ & $\chi^{2}$ \\
 & &    (ms)     &            &     (ms)    &            &     (ms)    &            &     (ms)    &            &     (ms)    &            &     (ms)    &         &     (ms)    &     &     (ms)    &         \\
}

\startdata
1 & J0332+5434 & 50$\pm$5 & 1.0 & 29$\pm$2 & 0.7 & 21$\pm$1 & 0.8 & 13.8$\pm$0.5 & 1.0 & 9.3$\pm$0.4 & 0.9 & 5.7$\pm$0.2 & 1.0 & 4.0$\pm$0.1 & 1.0 & 3.4$\pm$0.2 & 0.8 & - & - \\
2 & J1825-0935 & - & - & - & - & 35$\pm$4 & 1.4 & 20$\pm$2 & 0.9 & 11$\pm$1 & 0.7 & 9$\pm$2 & 9 & 6$\pm$1 & 0.9 & - & - & - & -\\
3 & J2219+4754 & - & - & 83$\pm$10 & 0.9 & 64$\pm$5 & 1.0 & 37$\pm$2 & 0.9 & 28$\pm$2 & 1.0 & 19$\pm$1 & 0.8 & 14$\pm$1 & 0.8 & 9.2$\pm$0.4 & 0.9 & 7.0$\pm$0.4 & 0.9 \\
\enddata

\end{deluxetable}

\clearpage

\startlongtable
\begin{deluxetable}{cllcccccclcl}

\tablecolumns{12}
\tablewidth{0pt}  
\tabletypesize{\scriptsize}
\tablecaption{Parameters of pulsars in our sample along with our measurements. For each pulsar, the table
lists it's name in J2000, position in the Galactic longitude and latitude, dispersion measure, distance from
ATNF pulsar catalogue \citep{mht05}, pulse width of the high frequency template used (W10 of the high
frequency profile), frequency scaling index $\alpha$, $\tau_{sc}$ scaled to 1 GHz with error bars estimated 
from the limits on $\alpha$, log $C_{n_{e}}^{2}$ and $\tau_{sc}$ from NE2001 model \citep{cl02}. Pulsars names 
with $\gamma$ as superscript are the ones which have previous $\alpha$ measurements.
\label{tab3}}
\tablehead{
\colhead{No.}  & \colhead{PSR} &  \colhead{l} & \colhead{b} &  \colhead{Period} & \colhead{DM} & \colhead{Distance} & \colhead{Width} & \colhead{$\alpha$} & \colhead{$\tau_{sc}$}(1 GHz) & \colhead{log $C_{n_{e}}^{2}$} & $\tau_{sc}$ (NE2001) \\
\colhead{} & \colhead{} &  \colhead{(deg)} & \colhead{(deg)} & \colhead{(s)} & \colhead{(pc cm$^{-3}$)} & \colhead{(kpc)} & \colhead{(ms)} & \colhead{} & \colhead{(ms)} & \colhead{m$^{-20/3}$} & (ms) \\
}
\startdata
1   &  J0332+5434$^{\gamma}$	&	144.995	&	-1.221	&	0.714555	&	26.76	&	1.00	&	7.3 & 4.3$\pm$0.1		& 6.1E$-$5 $\pm$ 0.4E$-$5	&	-3.92	&	7E$-$5	\\
2   &  J0502+4654		&	160.363	&	3.077	&	0.638542	&	42.19	&	1.79	&	15.0 & 2.1$^{+0.4}_{-0.7}$	& 0.7$^{+0.4}_{-0.2}$	&	-0.34	&	0.0003	\\
3   &  J0534+2200$^{\gamma}$	&	184.558	&	-5.784	&	0.033713	&	56.79	&	1.32	&	0.1 & 3.4$\pm$0.2		& 0.016 $\pm$ 0.002	&	-2.88	&	0.002	\\
4   &  J0614+2229$^{\gamma}$	&	188.785	&	2.400	&	0.334990	&	96.91	&	1.74	&	6.3 & 2.9$\pm$0.1		& 0.065 $\pm$ 0.004	&	-2.22	&	0.001	\\
5   &  J1328$-$4921		&	309.122	&	13.066	&	1.478829	&	118.00	&	8.40	&	7.5 & 4.0$^{+0.7}_{-0.6}$	& 0.03 $\pm$ 0.01	&	-5.84	&	0.003	\\
6   &  J1557$-$4258		&	335.273	&	7.952	&	0.329207	&	144.50	&	8.63	&	5.8 & 6.0$^{+0.4}_{-0.6}$	& 0.017$^{+0.004}_{-0.009}$	&	-6.27	&	0.065	\\
7   &  J1604$-$4909		&	332.152	&	2.442	&	0.327437	&	140.80	&	3.22	&	3.7 & 4.0$^{+0.6}_{-1.2}$	& 0.02$^{+0.02}_{-0.01}$	&	-4.52	&	0.14	\\
8   &  J1613$-$4714		&	334.573	&	2.835	&	0.382398	&	161.20	&	3.52	&	7.8 & 3.4$^{+0.6}_{-1.2}$	& 0.13$^{+0.04}_{-0.16}$	&	-2.97	&	0.06	\\
9   &  J1703$-$3241		&	351.786	&	5.387	&	1.211843	&	110.31	&	3.17	&	12.8 & 3.5$\pm$0.1		& 0.13 $\pm$ 0.01	&	-2.74	&	0.01	\\
10  &  J1705$-$3423		&	350.720	&	3.975	&	0.255408	&	146.36	&	3.84	&	11.8 & 3.3$\pm$0.1		& 0.9 $\pm$ 0.1	&	-1.45	&	0.2	\\
11  &  J1722$-$3207		&	354.561	&	2.525	&	0.477179	&	126.06	&	2.93	&	11.4 & 3.6$\pm$0.3		& 0.3 $\pm$ 0.1	&	-1.99	&	0.1	\\
12  &  J1741$-$3927		&	350.555	&	-4.749	&	0.512230	&	158.50	&	4.62	&	8.8 & 3.0$^{+0.1}_{-0.2}$	& 1.1$^{+0.1}_{-0.2}$	&	-1.69	&	0.02	\\
13  &  J1743$-$1351		&	12.699	&	8.205	&	0.405352	&	116.30	&	3.50	&	10.75 & 4.6$^{+0.4}_{-0.3}$	& 0.0017 $\pm$ 0.0004	&	-6.53	&	0.005	\\
14  &  J1745$-$3040		&	358.553	&	-0.963	&	0.367449	&	88.37	&	0.20	&	7.9 & 2.9$^{+0.4}_{-0.5}$	& 0.06$^{+0.02}_{-0.03}$	&	-1.73	&	0.2	\\
15  &  J1759$-$2205		&	7.472	&	0.810	&	0.460995	&	177.16	&	3.26	&	7.1 & 3.6$^{+0.7}_{-1.3}$	& 0.07$^{+0.03}_{-0.11}$	&	-3.29	&	0.13	\\
16  &  J1801$-$0357		&	23.596	&	9.257	&	0.921518	&	120.37	&	5.75	&	12.7 & 5.4$\pm$0.3		& 0.0012 $\pm$ 0.0002	&	-7.74	&	0.003	\\
17  &  J1807$-$0847		&	20.061	&	5.587	&	0.163732	&	112.38	&	1.50	&	3.8 & 2.7$^{+0.3}_{-0.4}$	& 0.3 $\pm$ 0.1	&	-0.75	&	0.03	\\
18  &  J1807$-$2715		&	3.843	&	-3.257	&	0.827806	&	312.98	&	5.01	&	15.5 & 4.7$^{+0.2}_{-0.3}$	& 0.34$^{+0.04}_{-0.08}$	&	-2.78	&	0.29	\\
19  &  J1808$-$0813		&	20.634	&	5.750	&	0.876068	&	151.27	&	3.97	&	29.0 & 5.4$\pm$0.3		& 0.03 $\pm$ 0.01	&	-4.35	&	0.01	\\
20  &  J1816$-$2650		&	5.219	&	-4.906	&	0.592899	&	128.12	&	3.59	&	14.8 & 2.5$^{+0.6}_{-1.2}$	& 0.4$^{+0.1}_{-0.5}$	&	-2.10	&	0.01	\\
21  &  J1823$-$0154		&	28.081	&	5.256	&	0.759793	&	135.87	&	5.28	&	9.5 & 4.1 $\pm$ 0.2		& 0.13 $\pm$ 0.02	&	-3.67	&	0.05	\\
22  &  J1825$-$0935		&	21.449	&	1.324	&	0.769026	&	19.38	&	0.30	&	12.0 & 5.0$\pm$0.5		& 1.2E$-$5 $\pm$ 0.3E$-$5	&	-2.48	&	0.0002	\\
23  &  J1835$-$1106		&	21.222	&	-1.512	&	0.165923	&	132.68	&	3.16	&	4.4 & 3.3$\pm$0.1		& 0.22 $\pm$ 0.02	&	-2.30	&	0.027	\\
24  &  J1849$-$0636		&	26.773	&	-2.497	&	1.451366	&	148.17	&	3.85	&	13.9 & 3.9$^{+0.3}_{-0.4}$	& 0.4 $\pm$ 0.1	&	-2.14	&	0.03	\\
25  &  J1854+1050  		&	42.887	&	4.223	&	0.573199	&	207.20	&	6.93	&	26.8 & 3.5$^{+0.2}_{-0.3}$	& 0.9$^{+0.1}_{-0.2}$	&	-2.60	&	0.01	\\
26  &  J1854$-$1421		&	20.456	&	-7.209	&	1.146606	&	130.40	&	6.91	&	21.9 & 2.6$\pm$0.2		& 0.16 $\pm$ 0.02	&	-3.99	&	0.01	\\
27  &  J1903$-$0632		&	28.479	&	-5.679	&	0.431891	&	195.61	&	5.37	&	9.3 & 3.7$\pm$0.1		& 0.44 $\pm$ 0.03	&	-2.69	&	0.14	\\
28  &  J1904$-$1224		&	23.291	&	-8.490	&	0.750811	&	118.23	&	7.27	&	6.8 & 5.4$^{+0.6}_{-1.0}$	& 0.04$^{+0.01}_{-0.04}$	&	-5.25	&	0.003	\\
29  &  J1905$-$0056		&	33.690	&	-3.551	&	0.643185	&	229.13	&	7.64	&	4.9 & 5.0$^{+0.6}_{-1.3}$	& 0.07$^{+0.02}_{-0.10}$	&	-4.87	&	0.09	\\
30  &  J1910$-$0309		&	32.280	&	-5.680	&	0.504606	&	205.53	&	6.07	&	8.6 & 4.2$^{+0.4}_{-0.6}$	& 0.014$^{+0.003}_{-0.007}$	&	-5.79	&	0.009	\\
31  &  J1916+1312$^{\gamma}$	&	47.576	&	0.451	&	0.281843	&	237.01	&	4.50	&	5.7 & 3.2$^{+0.7}_{-1.4}$	& 0.1$^{+0.1}_{-0.2}$	&	-3.45	&	0.1	\\
32  &  J1926+0431		&	40.980	&	-5.674	&	1.074071	&	102.24	&	4.99	&	15.8 & 2.4$^{+0.3}_{-0.2}$	& 0.20$^{+0.04}_{-0.03}$	&	-3.21	&	0.001	\\
33  &  J1932+2020$^{\gamma}$	&	55.575	&	0.639	&	0.268217	&	211.15	&	5.00	&	5.7 & 3.3 $\pm$ 0.2		& 0.7 $\pm$ 0.1	&	-2.20	&	0.14	\\
34  &  J1935+1616$^{\gamma}$	&	52.436	&	-2.093	&	0.358739	&	158.52	&	3.70	&	7.2 & 3.1$\pm$0.1		& 0.048 $\pm$ 0.003	&	-3.82	&	0.008	\\
35  &  J2004+3137$^{\gamma}$	&	69.011	&	0.021	&	2.111256	&	234.82	&	8.00	&	19.8 & 4.0$^{+0.4}_{-0.7}$	& 0.07$^{+0.02}_{-0.04}$	&	-4.96	&	0.06	\\
36  &  J2029+3744		&	76.898	&	-0.727	&	1.216771	&	190.66	&	5.77	&	20.1 & 4.1$^{+0.3}_{-0.4}$	& 0.12$^{+0.02}_{-0.04}$	&	-3.98	&	0.01	\\
37  &  J2055+3630$^{\gamma}$	&	79.133	&	-5.589	&	0.221499	&	97.31	&	5.00	&	3.7 & 3.4$^{+0.4}_{-0.7}$	& 0.16$^{+0.04}_{-0.10}$	&	-3.41	&	0.04	\\
38  &  J2113+4644		&	89.003	&	-1.266	&	1.014642	&	141.26	&	4.00	&	32.0 & 4.2 $\pm$ 0.1		& 0.011$\pm$ 0.001	&	-5.23	&	0.004	\\
39  &  J2219+4754		&	98.385	&	-7.598	&	0.538440	&	43.50	&	2.39	&	7.6 & 4.2$\pm$0.1		& 2.0E$-$4 $\pm$ 0.1E$-$4	&	-4.79	&	0.0001	\\
\enddata

\end{deluxetable}

\end{document}